\documentclass[runningheads]{llncs}

\usepackage{graphicx}
\usepackage[bookmarksopen=true]{hyperref}
\usepackage{cleveref}
\usepackage[frozencache,cachedir=minted-cache]{minted}
\usepackage[labelfont=bf]{caption}

\begin{document}
\setcounter{tocdepth}{5}
\title{Developing a High-Performance Process Mining Library with Java and Python Bindings in Rust}
\titlerunning{High-Performance Process Mining Library in Rust}
\author{Aaron Küsters\orcidID{0009-0006-9195-5380}, 
Wil M.P. van der Aalst\orcidID{0000-0002-0955-6940}}
\authorrunning{Aaron Küsters, Wil van der Aalst}

\institute{Process and Data Science (PADS), RWTH Aachen University, Germany 
\email{\{kuesters,wvdaalst\}@pads.rwth-aachen.de}}
\maketitle              % typeset the header of the contribution
\begin{abstract}
The most commonly used open-source process mining software tools today are ProM and PM4Py, written in Java and Python, respectively.
Such high-level, often interpreted, programming languages trade off performance with memory safety and ease-of-use. In contrast, traditional compiled languages, like C or C++, can achieve top performance but often suffer from instability related to unsafe memory management.
Lately, Rust emerged as a highly performant, compiled programming language with inherent memory safety.
In this paper, we describe our approach to developing a shared process mining library in Rust with bindings to both Java and Python, allowing full integration into the existing ecosystems, like ProM and PM4Py. 
By facilitating interoperability, our methodology enables researchers or industry to develop novel algorithms in Rust once and make them accessible to the entire community while also achieving superior performance.

\end{abstract}
\section{Introduction}
\label{sec:intro}
The field of \emph{process mining} is concerned with analyzing the execution of (business) processes.
Most techniques situated in the field leverage event data of past process executions to gain insights into processes.
Process mining techniques can mostly be categorized into the subfields of \emph{process discovery}, \emph{conformance checking} and \emph{performance analysis}.
\emph{Process discovery} is concerned with discovering process models, often Petri nets, from event data.
Based on a given process model and event data, \emph{conformance checking} can be applied to measure how well the past executions conform to the restrictions of the process model.
Similarly, \emph{performance analysis} can identify performance issues in actual process executions, like bottlenecks, based on a process model and event data.

There are dozens of commercial process mining tools, like \emph{Celonis}, \emph{SAP Signavio} or \emph{UiPath Process Mining}.
As for open-source solutions, \emph{ProM} and \emph{PM4Py} are the two most popular process mining tools today.
The ProM framework was first developed in 2005 and features a graphical user interface (GUI)~\cite{prom-new-era-2005}.
ProM is based on a plugin system, which allows researchers to easily make newly developed techniques available to the public.
Users can download and update plugins using an included plugin manager.
The ProM framework and its plugins are implemented in Java.
PM4Py (Process Mining for Python) is a Python library first presented in 2019~\cite{pm4py-bridging-the-gap-2019}.
PM4Py has no plugin system or GUI, but instead exposes its own implementations of popular techniques and algorithms as a Python software library.
This library can, in turn, be used by researchers and other end users for applying process mining directly or developing their own techniques as Python programs.

Some very popular and established process mining techniques are implemented in both PM4Py and ProM, however novel, lesser-known or specialized techniques are often only implemented for one of the two.
Porting implementations across the two solutions, and thus also across Java and Python, involves considerable effort and care, also caused by the different associated programming paradigms.
This situation creates a fragmentation of \emph{where} novel algorithms are implemented.
Researchers or industry typically cannot afford or justify the investment in time and money to develop their approach with both ProM and PM4Py, and instead are forced to choose just one option.
This fragmentation decreases the immediate usability and flexibility of newly developed approaches, which is not only disadvantageous for their creator because of the smaller potential audience but also hinders innovation (e.g., new scientific publications building on top of past work).

As mentioned previously, ProM is implemented in the Java programming language, while PM4Py is implemented in Python.
Java and Python are both popular high-level, often at least partially interpreted, languages with automatic garbage collection.
In particular, this means that programs written in those languages are commonly not directly compiled to machine code but instead transformed to an intermediate lower-level representation, which is translated to machine code when actually running the program by an associated interpreter.
Nowadays, these transformations are also combined with so-called \emph{just-in-time compilation}, leveraging information about the current execution for further optimization.
The interpreter also keeps track of the allocated memory and automatically manages memory allocation and deallocation when needed (\emph{garbage collection}).
These concepts make Java and Python easy to use and programs written in them memory safe (i.e., all references point to valid objects in memory), but not without cost.
Depending on the actual use case, there is an expected performance penalty associated with interpreted and garbage-collected language runtimes because of the corresponding overhead of these tasks.
Thus, based on these criteria, Java and Python trade off performance for memory safety and ease-of-use.
The programming languages \emph{C} or \emph{C++}, on the other hand, are situated on the other extreme of this trade-off: Memory management is designated as a responsibility of the programmer and the language is compiled to machine code, but it allows leveraging very high performance.
As memory references have to be managed by the programmer, programs written in C often suffer from problems related to unsafe memory management, like crashes, subtle bugs, or also security vulnerabilities.

Lately, the programming language \emph{Rust} has gained popularity as a highly performant, compiled language with a focus on memory safety and concurrency.
Rust guarantees memory safety using a so-called \emph{borrow checker}, which, as part of the compilation, analyzes and tracks the lifetime of references across the program.
This approach guarantees full memory safety without a garbage collector and the associated performance penalty.

\begin{figure}
    \centering
    \def\svgwidth{\columnwidth}
    %% Creator: Inkscape 1.3.2 (091e20ef0f, 2023-11-25, custom), www.inkscape.org
%% PDF/EPS/PS + LaTeX output extension by Johan Engelen, 2010
%% Accompanies image file 'test.pdf' (pdf, eps, ps)
%%
%% To include the image in your LaTeX document, write
%%   \input{<filename>.pdf_tex}
%%  instead of
%%   \includegraphics{<filename>.pdf}
%% To scale the image, write
%%   \def\svgwidth{<desired width>}
%%   \input{<filename>.pdf_tex}
%%  instead of
%%   \includegraphics[width=<desired width>]{<filename>.pdf}
%%
%% Images with a different path to the parent latex file can
%% be accessed with the `import' package (which may need to be
%% installed) using
%%   \usepackage{import}
%% in the preamble, and then including the image with
%%   \import{<path to file>}{<filename>.pdf_tex}
%% Alternatively, one can specify
%%   \graphicspath{{<path to file>/}}
%% 
%% For more information, please see info/svg-inkscape on CTAN:
%%   http://tug.ctan.org/tex-archive/info/svg-inkscape
%%
\begingroup%
  \makeatletter%
  \providecommand\color[2][]{%
    \errmessage{(Inkscape) Color is used for the text in Inkscape, but the package 'color.sty' is not loaded}%
    \renewcommand\color[2][]{}%
  }%
  \providecommand\transparent[1]{%
    \errmessage{(Inkscape) Transparency is used (non-zero) for the text in Inkscape, but the package 'transparent.sty' is not loaded}%
    \renewcommand\transparent[1]{}%
  }%
  \providecommand\rotatebox[2]{#2}%
  \newcommand*\fsize{\dimexpr\f@size pt\relax}%
  \newcommand*\lineheight[1]{\fontsize{\fsize}{#1\fsize}\selectfont}%
  \ifx\svgwidth\undefined%
    \setlength{\unitlength}{571.5bp}%
    \ifx\svgscale\undefined%
      \relax%
    \else%
      \setlength{\unitlength}{\unitlength * \real{\svgscale}}%
    \fi%
  \else%
    \setlength{\unitlength}{\svgwidth}%
  \fi%
  \global\let\svgwidth\undefined%
  \global\let\svgscale\undefined%
  \makeatother%
  \begin{picture}(1,0.49889764)%
    \lineheight{1}%
    \setlength\tabcolsep{0pt}%
    \put(0,0){\includegraphics[width=\unitlength,page=1]{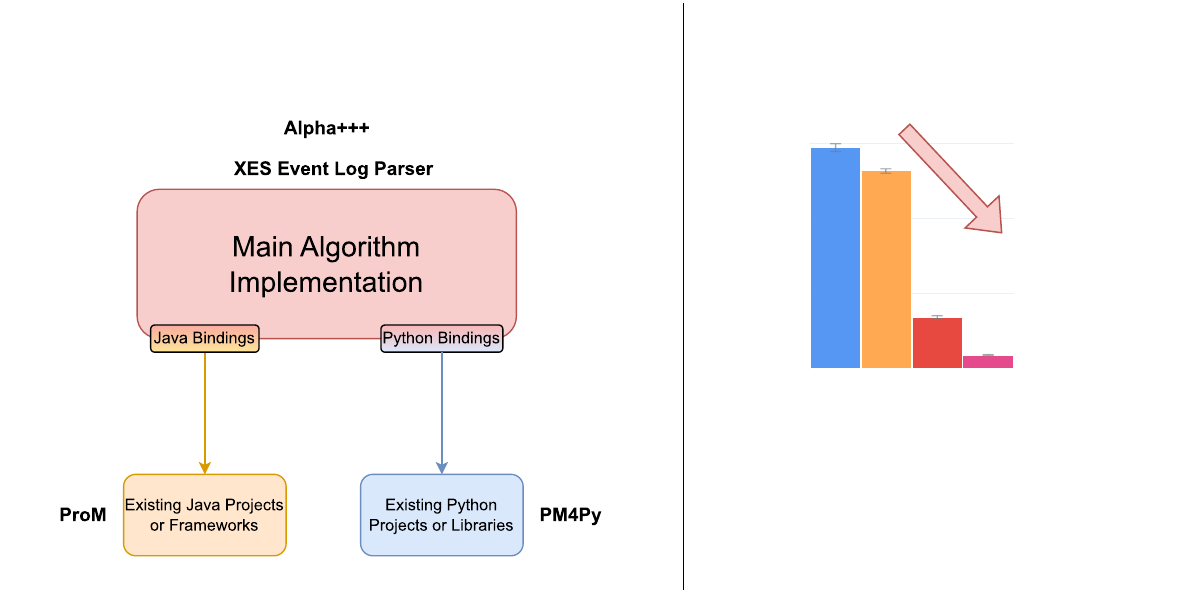}}%
    \put(0.15425007,0.4265773){\makebox(0,0)[lt]{\lineheight{1.25}\smash{\begin{tabular}[t]{l}See \color{blue}\nameref{sec:implementation}\end{tabular}}}}%
    \put(0.0434166,0.46392236){\makebox(0,0)[lt]{\lineheight{1.25}\smash{\begin{tabular}[t]{l}\textbf{Implement Once, Use Everywhere}\end{tabular}}}}%
    \put(0.67937937,0.43034959){\makebox(0,0)[lt]{\lineheight{1.25}\smash{\begin{tabular}[t]{l}See \color{blue}\nameref{sec:evaluation}\end{tabular}}}}%
    \put(0.65542299,0.46769465){\makebox(0,0)[lt]{\lineheight{1.25}\smash{\begin{tabular}[t]{l}\textbf{Improved Speed}\end{tabular}}}}%
    \put(0,0){\includegraphics[width=\unitlength,page=2]{figures/introduction.pdf}}%
  \end{picture}%
\endgroup%

    \caption{The two main advantages of our proposed approach.
    The main algorithm (like Alpha+++ or XES parsing in this paper) is implemented only once in Rust. Java and Python bindings make this implementation available to established tool ecosystems (like ProM and PM4Py).
    Additionally, our evaluation indicates great potential speedups of Rust implementations, compared to baseline implementations in Java or Python.}
    \label{fig:introduction}
\end{figure}

In this paper, we present an approach of implementing algorithms once in Rust and then creating language bindings (e.g., for Java and Python) to make this implementation accessible from other programming languages and environments.
While the algorithms and integrations implemented in the context of this paper are situated in the field of process mining, the approach is generally applicable.
\autoref{fig:introduction} shows an overview of the main advantages of the proposed implementation approach.

The remainder of this paper is organized as follows.
First, we discuss related work in \Cref{sec:related-work}.
We continue with elaborating on our implementation approach in \Cref{sec:implementation}, covering the overall architecture, implementation details and our provided starter kit template.
Next, in \Cref{sec:evaluation}, we evaluate the runtime of our implementations by measuring and comparing the execution time of both the Alpha+++ process discovery algorithm and an XES event log parser.
Finally, we conclude this paper in \Cref{sec:conclusion}.

\section{Related Work}
\label{sec:related-work}
In this section, we first present related work on the example process discovery algorithm used throughout the later sections (\Cref{sec:implementation} and \Cref{sec:evaluation}).
Next, we also explore prior scientific work on Rust implementations and runtime performance comparisons across programming languages.

There are many different types of process discovery techniques and concrete algorithms, like the Inductive miner or Alpha algorithm.
These algorithms aim to construct a process model, commonly in the form of an accepting Petri net, based on an input event log.
In the later sections, we use the Alpha+++ algorithm, which is based on the classic Alpha algorithm, as an example implementation.
For a more detailed overview of process discovery, we refer interested readers to \cite{handbook-foundations-of-discovery,process-mining-in-action--discovery}.
The Alpha+++ algorithm presented in \cite{alpha-revisit-2023} is a process discovery algorithm with a focus on real-life event data.
It builds on top of the foundational concepts of the original Alpha algorithm, which was introduced in \cite{alpha-algorithm-workflow-mining-2004}.
The result of these algorithms is a process model in the form of an accepting Petri net.
In particular, these algorithms construct a set of Petri net places by first extracting directly follows relationships of the input event log.
The Alpha+++ algorithm has multiple consecutive filtering steps to remove place candidates which do not fit to the observed behavior in the input event log.
First, aggregated frequency information is used to filter out clearly unbalanced place candidates.
This can be done quickly because it does not require iterating over the full input event log for every place candidate.
As a second filtering step, the remaining place candidates are additionally filtered by replaying the observed behavior in the log.
This allows filtering even more candidates, but is computationally expensive.

There are a few published case studies of researchers or industry using Rust for implementation projects.
In~\cite{rust-bio2015}~Kösters presents a bioinformatics library written in Rust, focusing on speed and memory safety.
The author compared the performance of the newly implemented algorithms with established C++-based implementations and observed comparable results.
In~\cite{rust-astrophysics-2016}, the authors explore the potential advantages of using Rust in the context of astrophysics, where programming languages like Fortran or C are largely used.
As a proof of concept, the authors implemented a simulation algorithm in Rust, Fortran, C and Go and compared the execution times.
In their evaluation, Rust achieved the highest performance, closely followed by Fortran.
Similarly, in~\cite{performance-vs-programming-effort-2021},~Costanzo et al.~investigated both the performance and required programming effort for algorithms in the context of High-Performance Computing (HPC).
The performance evaluation indicated that, while C has some smaller performance advantage in certain configurations, the overall performance is largely comparable between Rust and C.
In the evaluation of programming effort, however, Rust proved advantageous because of the provided high-level language features, easy parallelization and memory safety.

\section{Implementation}
\label{sec:implementation}
% Details on Implementation (General, Implementation of Java and Python bindings, Different ways to pass data, \dots)
In this section, we describe the implementation details of our approach.
We start with an architectural overview.
Next, we discuss details on FFI bindings to Java and Python.
After that, we present WebAssembly as an alternative binding method and demonstrate the possibility of JavaScript bindings. 
Subsequently, we discuss methods of exchanging data between the main implementation and the thin language wrappers.
Next, we present an overview of porting the Alpha+++ algorithm to Rust and implementing the required process mining functionality, as well as bindings to Python and Java.
Finally, we provide a starting kit template of the proposed approach for interested readers.

\subsection{Overview}
\begin{figure}[h!]
    \centering  
    \includegraphics[width=0.6\textwidth]{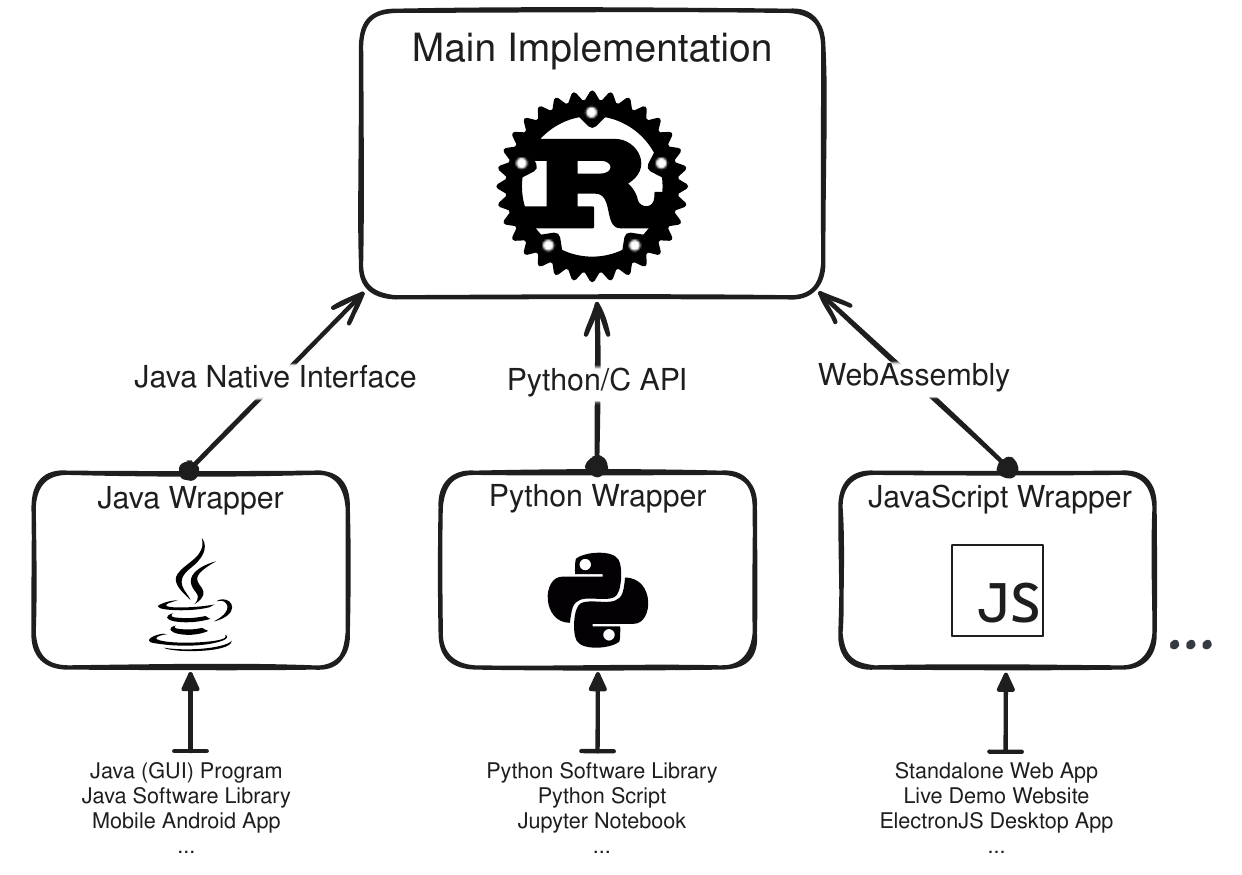}
    \caption[Overview of the approach]{Overview of the approach: A main implementation is written once in Rust. Thin wrappers for Java, Python, or other languages bind to the main implementation and expose functionality for easy use.
    Other programs (like a Java GUI program, a Python script) can make use of the exposed functionality.}
    \label{fig:overview}
\end{figure}

The main idea of our approach is, simply put, \emph{write once, use everywhere}.
\autoref{fig:overview} shows an architectural overview of the different implementation components.
Complex algorithms and logic is (once) implemented in a Rust software library.
To make this implementation accessible from other languages, platforms and programs, \emph{thin wrappers} written in the target languages are used.
They are \emph{thin}, as they only serve as a simple interface to the main Rust implementation.
In particular, they do not require duplicate implementation of the complex algorithms or logic.
Thus, not much implementation effort is required in exposing the main program to multiple other frameworks, programs, and languages.
The complete shared library implementation is available as an open-source project.\footnote{\url{https://github.com/aarkue/rust-bridge-process-mining/}}

\subsection{FFI Bindings to Java and Python}
In this subsection, we will take a closer look into how bindings can be implemented for Java and Python.
Both languages feature powerful and mature foreign function interfaces (FFI) that allow dynamically calling native functions from a compiled library.
For Java, this functionality is implemented as part of the \emph{Java Native Interface} (JNI).
For Python, it is exposed as the \emph{Python/C API}.
Most other popular languages also expose similar functionality.
In \autoref{code:basic-jni}, we showcase a very basic usage example of a native library function using Java's JNI.
For Python, the native functionality can be similarly exposed as a Python package with native implementation parts.
\begin{listing}[h]
    \centering
    \caption{An example Java class using the (imaginary) native library \texttt{native\_fib} to calculate numbers of the Fibonacci sequence.
    The native function \texttt{fibonacci} can only be used inside this class.
    For outside use, a different function \texttt{calcFibonacciNumber} is exposed, which performs additional argument checks.}
    \label{code:basic-jni}
\begin{minted}{java}
class FibHelper {
    static {
        System.loadLibrary("native_fib"); // Load native library
    }

    private static native int fibonacci(int n); // Declare native method

    public static int calcFibonacciNumber(int n) { // Expose functionality
        if (n < 0) {
            throw new IllegalArgumentException("Position must be >= 0");
        } else {
            return fibonacci(n);
        }
    }
}
\end{minted}
\end{listing}

It is generally recommended to define or use native functionality in as few places as possible~\cite{jni-guide}.
This is also one of the reasons why we advise implementing thin wrappers in the target language (i.e., in Java and Python) to handle the native library calls.
Additionally, wrappers can implement error handling or argument checking and other safeguards, like the \texttt{calcFibonacciNumber} function in \autoref{code:basic-jni}.

So far, we primarily looked at the implementations of the thin wrapper bindings written in the target languages.
On the Rust side, we make use of the libraries \emph{PyO3}\footnote{\url{https://github.com/PyO3/pyo3}} and \emph{jni}\footnote{\url{https://github.com/jni-rs/jni-rs}}, to implement bindings to Python and Java, respectively.
Both of the libraries implement the core binding functionality and provide helpers to convert passed arguments to and from the target language.

Finally, to use the native Rust code from Java or Python, the Rust code is first compiled into a shared/dynamic library (i.e., \texttt{.so} or \texttt{.dll} files).
That library file is then loaded by the target language wrapper, as shown in the top part of \autoref{code:basic-jni} for Java.
For Python, the underlying concepts are the same, but \emph{Maturin}\footnote{\url{https://github.com/PyO3/maturin}} automatically handles building Python libraries with these bindings.

\subsection{WebAssembly and JavaScript}
In addition to the previously presented Java and Python binding options, we also introduce WebAssembly as a potential additional compilation target.
While there are also FFI bindings for the Node.js JavaScript runtime, JavaScript execution in the browser cannot make use of such features.
Instead, there exists another approach to allow high-performance programs to run in web browsers and other platforms.
\emph{WebAssembly} is a portable binary program format that can nowadays be executed in all major browsers.
Rust programs or libraries can easily be compiled to WebAssembly, although a few restrictions apply.
For example, some features of the Rust standard library (e.g., file access \texttt{std::io} or system time \texttt{std::time}) cannot be used for general WebAssembly targets.\footnote{There are efforts to standardize a system interface for WebAssembly to also provide such additional features. See also \url{https://wasi.dev/}.}
However, general computing tasks and many of the popular available Rust libraries work flawlessly when executed as WebAssembly. 

WebAssembly provides full portability, which allows users to try out WebAssembly programs without downloading, configuring or installing any additional software.
Apart from that, WebAssembly also provides sandboxed execution.
This makes WebAssembly not only an attractive format for targeting browsers and the web, but also to implement portable, sandboxed, and high-performance programs in general.

\subsection{Exchanging Data}
In this subsection, we will describe different approaches to passing input data and arguments from and to functions of the shared Rust implementation.
The data exchange techniques presented here all make an independent copy of the passed data.
This allows fast execution without any communication overhead after the initial transfer of data.
As an alternative, it would also be possible to interact with the Java or Python execution environment from Rust, e.g., to evaluate function calls or object properties.
We argue, however, that copying all required data once instead creates a clearer separation of concerns and makes the shared implementation more easily re-useable and also easier to implement.

\subsubsection{Basic data types}
Simple function arguments, like strings, integer or floating-point numbers, can be passed with little to no additional work.
For the Python bindings using PyO3 all commonly used types can be automatically converted from the Python type to the corresponding Rust type.
See \autoref{code:basic-arguments-pyfunction} for an example.
For the Java bindings, there are similar automatic conversion for the primitive data types in Java (e.g., \texttt{int} or \texttt{double}).
However, strings passed to and from Java need to be explicitly converted, as shown in \autoref{code:basic-arguments-jni}.

\begin{listing}[p]
    \centering
    \caption{Example Rust function with Python bindings. PyO3 implements the traits \texttt{FromPyObject} and \texttt{IntoPy<PyObject>} for many existing types.
    This enables implicit, automatic conversion of the function arguments (\texttt{String}) and return value (\texttt{usize}).}
    \label{code:basic-arguments-pyfunction}
\begin{minted}{rust}
use pyo3::prelude::*;

#[pyfunction]
fn get_edit_distance(a: String, b: String) -> i32 {
    let dis = calc_edit_distance(&a,&b);
    return dis;
}
\end{minted}
\end{listing}

\begin{listing}[p]
    \centering
    \caption{Example Rust function with Java bindings. The jni library enables conversion from and to basic data types like \texttt{jint}/\texttt{i32}.
    \texttt{String}, however, requires explicit conversion.
    }
    \label{code:basic-arguments-jni}
\begin{minted}{rust}
use jni::{
    objects::{JClass, JString},
    sys::jint,
    JNIEnv,
};
use jni_fn::jni_fn;

#[jni_fn("org.example.EditDistance")]
pub fn getEditDistance<'local>(
    mut env: JNIEnv<'local>,
    _: JClass,
    a_jstr: JString,
    b_jstr: JString,
) -> jint {
    let a = &env.get_string(&a_jstr).unwrap().to_str().unwrap();
    let b = &env.get_string(&b_jstr).unwrap().to_str().unwrap();
    let dis = calc_edit_distance(&a, &b);
    return dis;
}
\end{minted}
\end{listing}

\subsubsection{Complex data structures}
More complex data structures can be serialized (for instance as JSON) and sent to the Rust side (or back) via different methods.
For example, the data could be encoded as a string, a byte array or a reference to a temporary file containing the data.
Building an object representation of the complex data structure on both involved execution sides enables easier use and manipulation of the data structures, leading to an improved developer experience.

\subsubsection{Persisting data in Rust}
If data should persist on the Rust side (e.g., because data only needs to be loaded once and multiple commands should be handled after that) we can break out of Rust's automatic reference counting by using Boxed/Unboxed values.
Fundamentally, this involves first loading the data in Rust and then returning a reference (long) to the place in memory where the data is stored back to the calling program.
On subsequent requests to the native Rust implementation, the reference is passed as long as well.
Rust can then unbox the data stored at this reference again and process it.
Finally, once the data can be unloaded, a native call to the Rust library can instruct the data to be unloaded.
Note, that while this can improve performance significantly for multiple requests involving the same data, it compromises some of the safety guarantees of Rust.
In particular, the caller in the target language must guarantee that the passed pointer is valid and cleaned up (by calling the corresponding unload function) eventually.

\subsection{Implementing Alpha+++ and Process Mining Basics}
\begin{figure}[h]
    \centering
    \includegraphics[width=\textwidth]{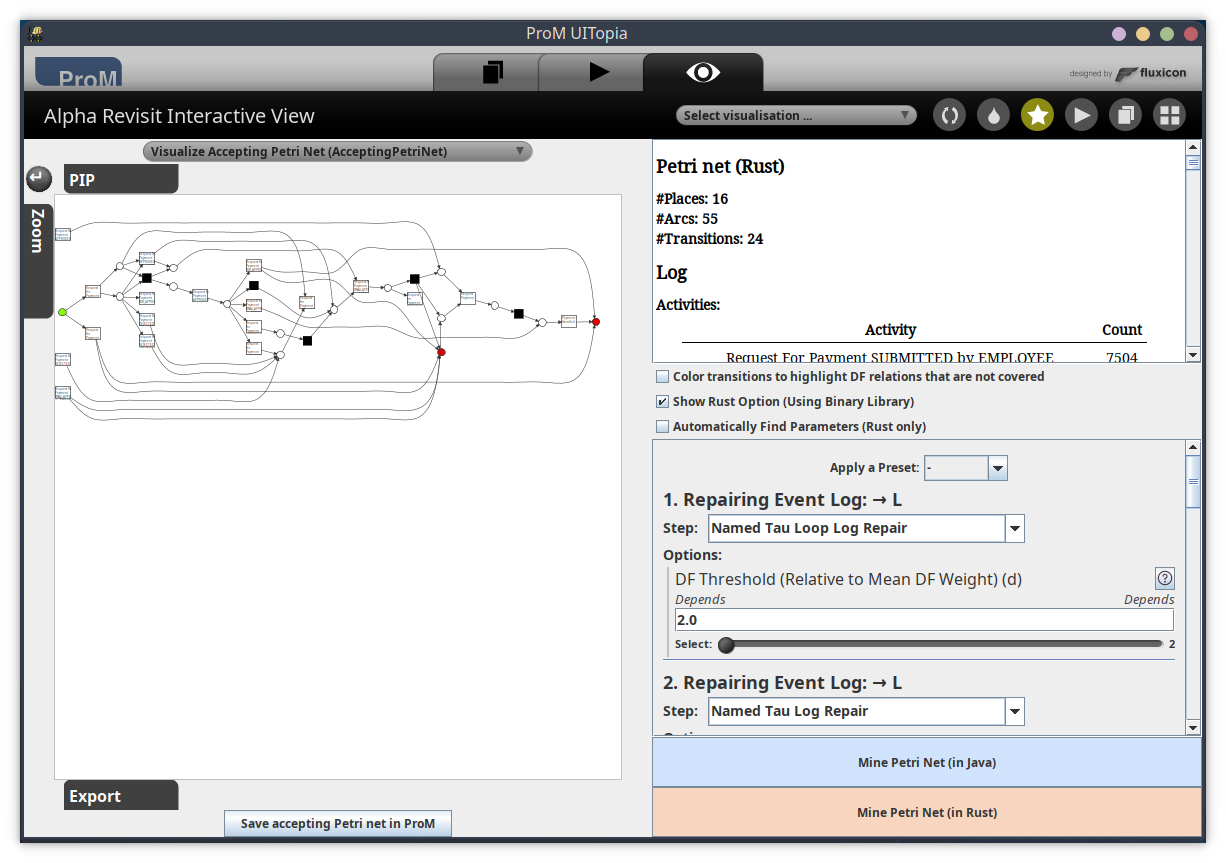}
    \caption{Screenshot of the AlphaRevisitExperiments ProM plugin featuring a \emph{Mine Petri Net (in Rust)} button for executing the Alpha+++ discovery in Rust instead of Java.}
    \label{fig:prom-mine-with-rust}
\end{figure}

We first implemented data structures and functionality for \emph{event logs}, \emph{directly-follows graphs} and \emph{Petri nets}, as a fundament for porting the Alpha+++ algorithm.
Similar to the previous Alpha+++ implementations in Python and Java, the Rust implementation utilizes the \emph{activity projection} of event logs, disregarding any additional event log information not needed for discovery.
Apart from saving memory, this approach also allows for quicker exchange of data from the wrapper implementations, as only a small part of the event log data needs to be transferred.
Our Alpha+++ Rust implementation is exposed in the AlphaRevisitExperiments\footnote{\url{https://github.com/promworkbench/AlphaRevisitExperiments}} ProM plugin using Java bindings, as shown in \autoref{fig:prom-mine-with-rust}.

\subsection{Starter Kit}
We published a starter kit template to make our implementation approach easily accessible.
It contains all the basic project structure and organization required to start developing a shared Rust library with Java and Python bindings.
The starter kit contains four main parts: The main shared library, Java bindings, Python bindings and a (executable) binary Rust project.
It also provides examples demonstrating how to consume the bindings in Java and Python code.
Interested readers can visit \url{https://github.com/aarkue/rust-bridge-template} to bootstrap their own project or experiment.
\clearpage
\section{Evaluation}
\label{sec:evaluation}
In this section, we will present results on the performance of the featured implementations.
We will start with a performance comparison of our implemented Rust-based XES event log importer and then continue with evaluating the implementations of the Alpha+++ algorithm.

All evaluations were run on a laptop with an AMD Ryzen 9 5900HX CPU (8 cores/16 threads) and 32 GB of memory, and each measurement was computed $N=10$ times. We report statistical error in the form of standard deviation in all figures and evaluation tables.

\begin{table}[h]
    \caption{Overview of the event logs used for evaluation.}
    \label{tab:evaluation-logs}
    \centering
    \begin{tabular}{l|c|c|c|c|c}
        \textbf{Event Log}                                                                  & \textbf{\#Events} & \textbf{\#Activities} & \textbf{\#Cases} & \textbf{\#Variants} & \textbf{Reference}      \\ \hline
        RTFM                                                                                & 561,470           & 11                    & 150,370           & 231                 & \cite{rtfm_log}         \\ \hline
        Sepsis                                                                              & 15,214            & 16                    & 1,050             & 846                 & \cite{sepsis_log}       \\ \hline
        \begin{tabular}[c]{@{}l@{}}BPI Challenge 2019\\ (Sample of 3000 Cases)\end{tabular} & 18,972            & 34                    & 3,000             & 470                 & \cite{bpic2019_log}     \\ \hline
        \begin{tabular}[c]{@{}l@{}}BPI Challenge 2020\\ (Request for Payment)\end{tabular}  & 36,796            & 19                    & 6,886             & 89                  & \cite{bpic2020_rfp_log} \\ \hline
        \begin{tabular}[c]{@{}l@{}}BPI Challenge 2020\\ (Domestic Declaration)\end{tabular} & 56,437            & 17                    & 10,500            & 99                  & \cite{bpic2020_dd_log}
    \end{tabular}
\end{table}

\subsection{XES Import}
We implemented an XES event log importer in Rust.
To make it easily available, we created performant Python bindings using polars\footnote{\url{https://github.com/pola-rs/polars/}}, which handles transferring the event log DataFrame.
The XES importer is available as the standalone Python package \emph{rustxes}, which is also published on PyPi.\footnote{\url{https://github.com/aarkue/rustxes} \url{https://pypi.org/project/rustxes/}} and can be used to import XES event logs as a polars DataFrame.
If both the rustxes and PM4Py package are installed, the Rust XES importer can be used directly from PM4Py.\footnote{Using \mintinline{python3}|pm4py.read_xes(log_path, variant="rustxes")|}
This PM4Py integration was implemented by Alessandro Berti and also supports transferring event data to the GPU for further processing.\footnote{\url{https://github.com/pm4py/pm4py-core/blob/release/pm4py/objects/log/importer/xes/variants/rustxes.py}}

To evaluate the performance of our XES importer, we imported the five commonly used XES event logs introduced in \autoref{tab:evaluation-logs} from Python.
Some of these XES event log files are also compressed as a \texttt{.xes.gz} archive, which all evaluated implementations can support as well.
We then measured the execution time for importing the XES files with the following three different parsing implementations:

\begin{description}
    \item[PM4Py (iterparse)] is the current standard XES import algorithm in PM4Py, which supports many advanced XES features and is certified against the X1 XES standard certification.
    \item[PM4Py (line\_by\_line)] is a different XES import algorithm included in PM4Py, which is a more simple implementation, supporting only basic XES features.
    \item[rustxes] is our proposed implementation. It can parse nested attributes, log globals and other advanced XES features, but currently only exposes such advanced information as JSON-encoded attribute value strings to the Python bindings.
\end{description}

\begin{figure}[h]
    \centering
    \includegraphics[width=0.49\textwidth]{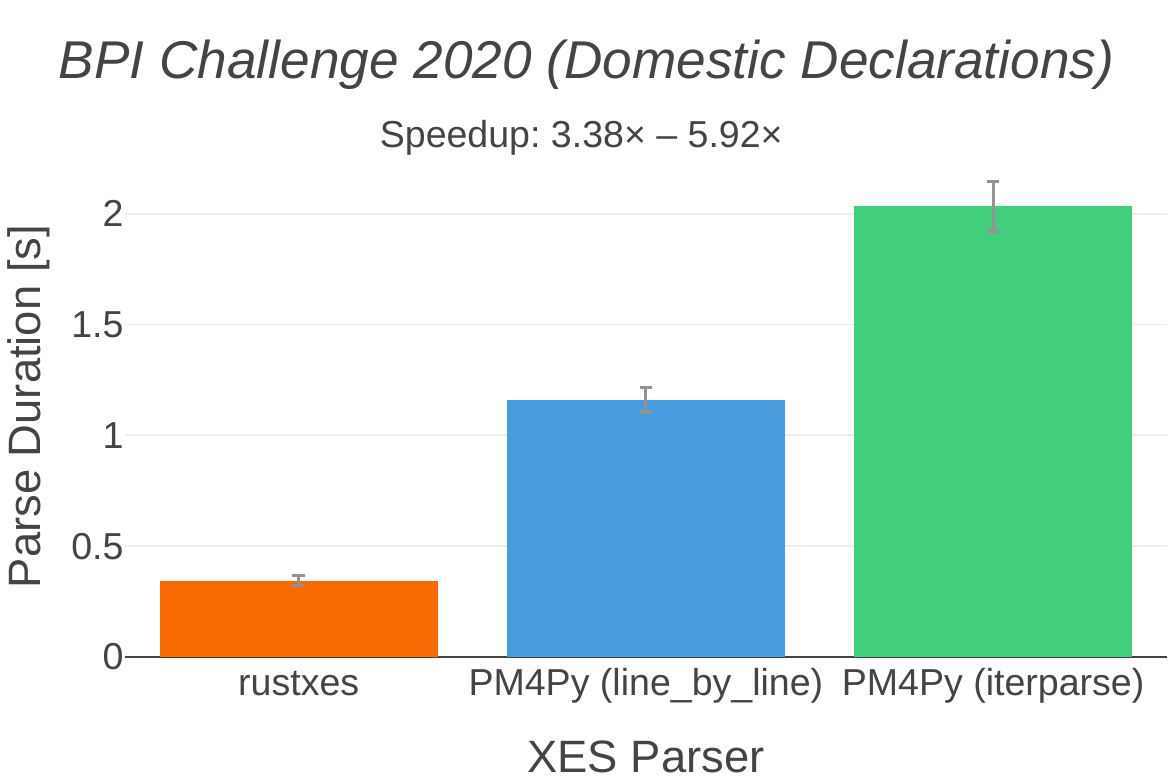}
    \includegraphics[width=0.49\textwidth]{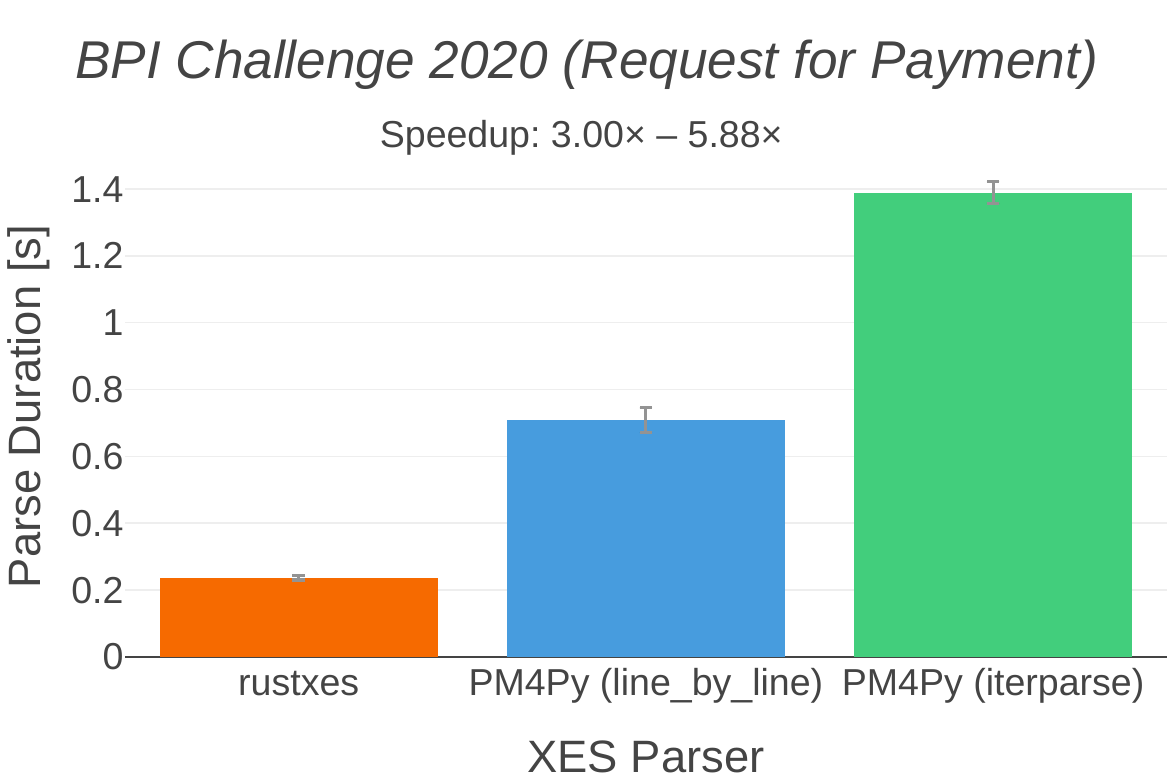}
    \includegraphics[width=0.49\textwidth]{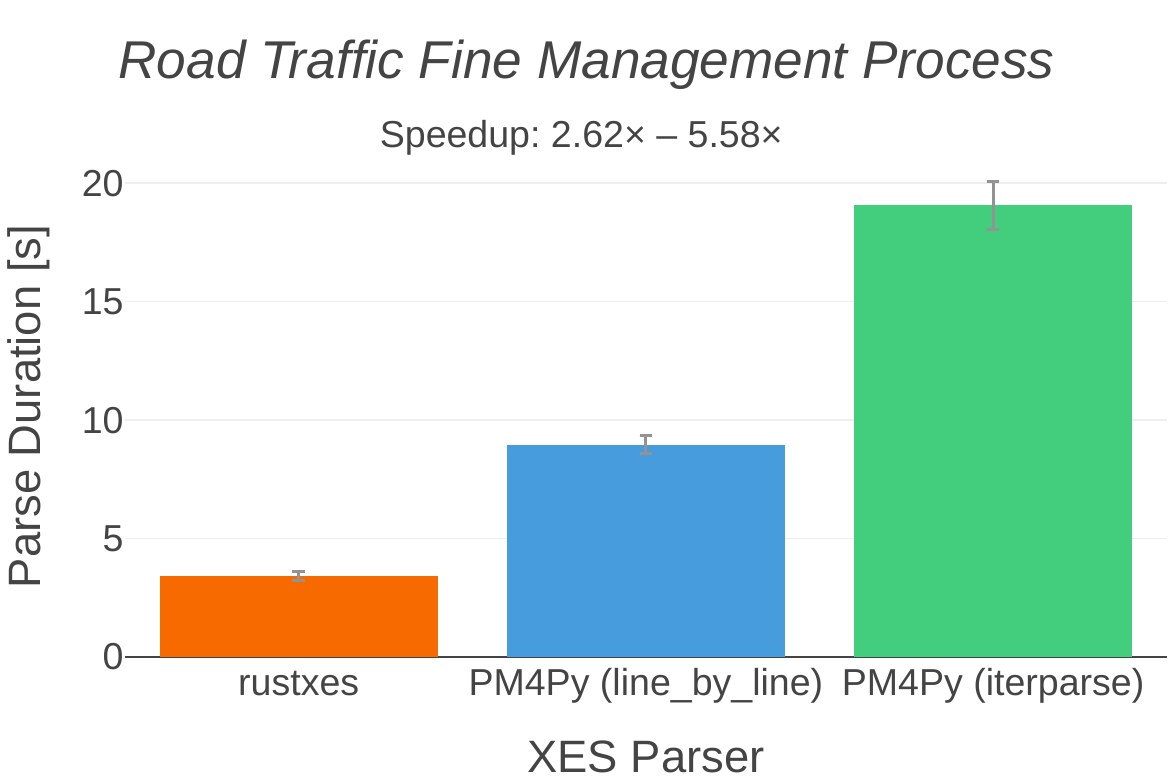}
    \includegraphics[width=0.49\textwidth]{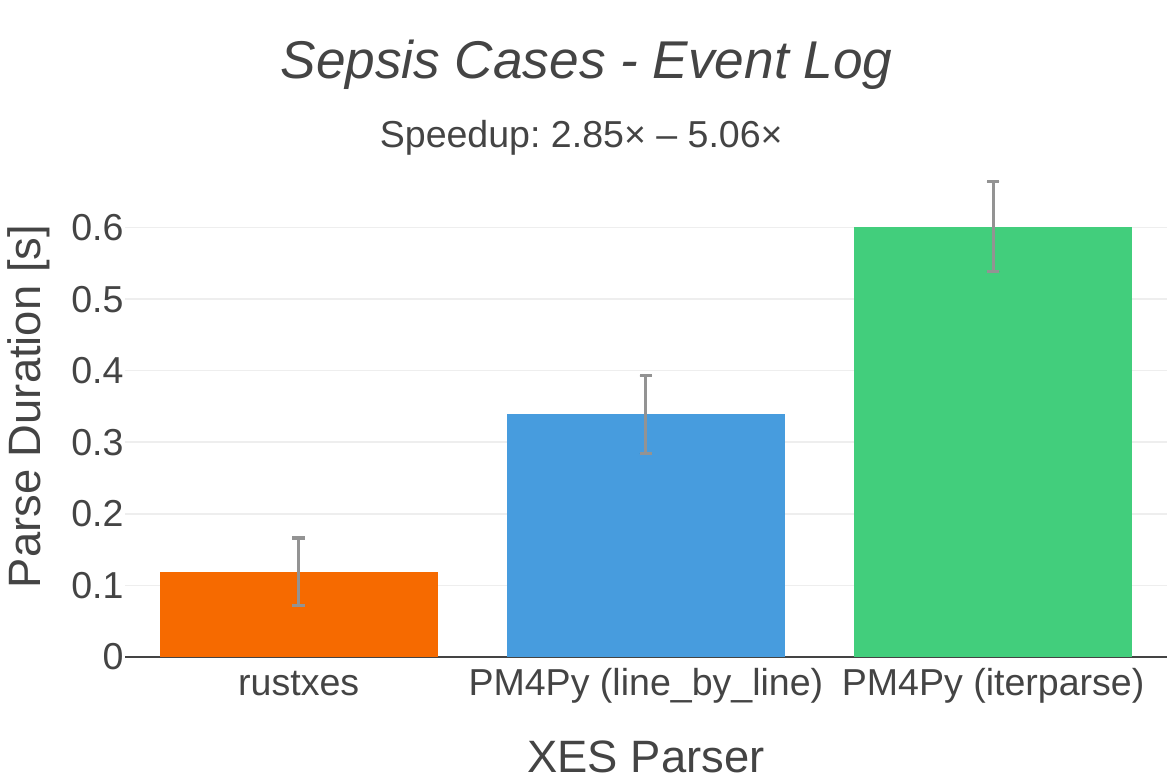}
    \caption{Performance comparison of XES import implementations. For all evaluated event logs, the rustxes performed better than the two other implementations. }
    \label{fig:xes-import-performance}
\end{figure}

In \autoref{fig:xes-import-performance}, we present some of our measured results.
Across the considered event logs, rustxes consistently outperforms the naive PM4Py (line\_by\_line) implementation by a factor of 2.5--3 and the more advanced PM4Py (iterparse) implementation by a factor of 5--6.
Results on other event logs, especially if they contain many instances of advanced XES features, may differ, as only rustxes and PM4Py (iterparse) actually parse such advanced information.
Note, that the reported durations for rustxes include all data transfer and data conversion from Rust to Python, and thus, in fact, demonstrate the potential speedup for a drop-in replacement.
\autoref{table:xes-parser-results} additionally shows all observed median, mean, and standard deviation values.

\setlength{\tabcolsep}{2pt}
\begin{table}[h]
    \centering
    \scriptsize
    \caption{Evaluation results for XES import implementations.}
    \label{table:xes-parser-results}
    \begin{tabular}{llcccll}
\hline
 Event Log                                                 & Parser                 &  Median  &   Mean   &   SD    \\
\hline
 \emph{BPI Challenge 2020}\\\emph{(Request for Payment)}   &                        &          &          &         \\
                                                           & rustxes                & 0.2365s  & 0.2364s  & 0.0063s \\
                                                           & PM4Py (line\_by\_line) & 0.7027s  & 0.7095s  & 0.0376s \\
                                                           & PM4Py (iterparse)      & 1.3958s  & 1.3889s  & 0.0335s \\
 \emph{BPI Challenge 2020}\\\emph{(Domestic Declarations)} &                        &          &          &         \\
                                                           & rustxes                & 0.3440s  & 0.3437s  & 0.0215s \\
                                                           & PM4Py (line\_by\_line) & 1.1402s  & 1.1612s  & 0.0561s \\
                                                           & PM4Py (iterparse)      & 2.0043s  & 2.0339s  & 0.1119s \\
 \emph{BPI Challenge 2019}\\\emph{(Sample of 3000 Cases)}  &                        &          &          &         \\
                                                           & rustxes                & 0.1196s  & 0.1332s  & 0.0241s \\
                                                           & PM4Py (line\_by\_line) & 0.3187s  & 0.3274s  & 0.0389s \\
                                                           & PM4Py (iterparse)      & 0.8274s  & 0.8283s  & 0.0433s \\
 \emph{Road Traffic Fine}\\\emph{Management Process}       &                        &          &          &         \\
                                                           & rustxes                & 3.4589s  & 3.4166s  & 0.1948s \\
                                                           & PM4Py (line\_by\_line) & 9.0920s  & 8.9453s  & 0.3792s \\
                                                           & PM4Py (iterparse)      & 19.0315s & 19.0525s & 1.0110s \\
 \emph{Sepsis Cases - Event Log}                           &                        &          &          &         \\
                                                           & rustxes                & 0.1016s  & 0.1189s  & 0.0472s \\
                                                           & PM4Py (line\_by\_line) & 0.3360s  & 0.3393s  & 0.0545s \\
                                                           & PM4Py (iterparse)      & 0.5857s  & 0.6014s  & 0.0632s \\
\hline
\end{tabular}
\end{table}

\subsection{Alpha+++}
\begin{figure}[p]
    \centering
    \includegraphics[width=0.95\textwidth]{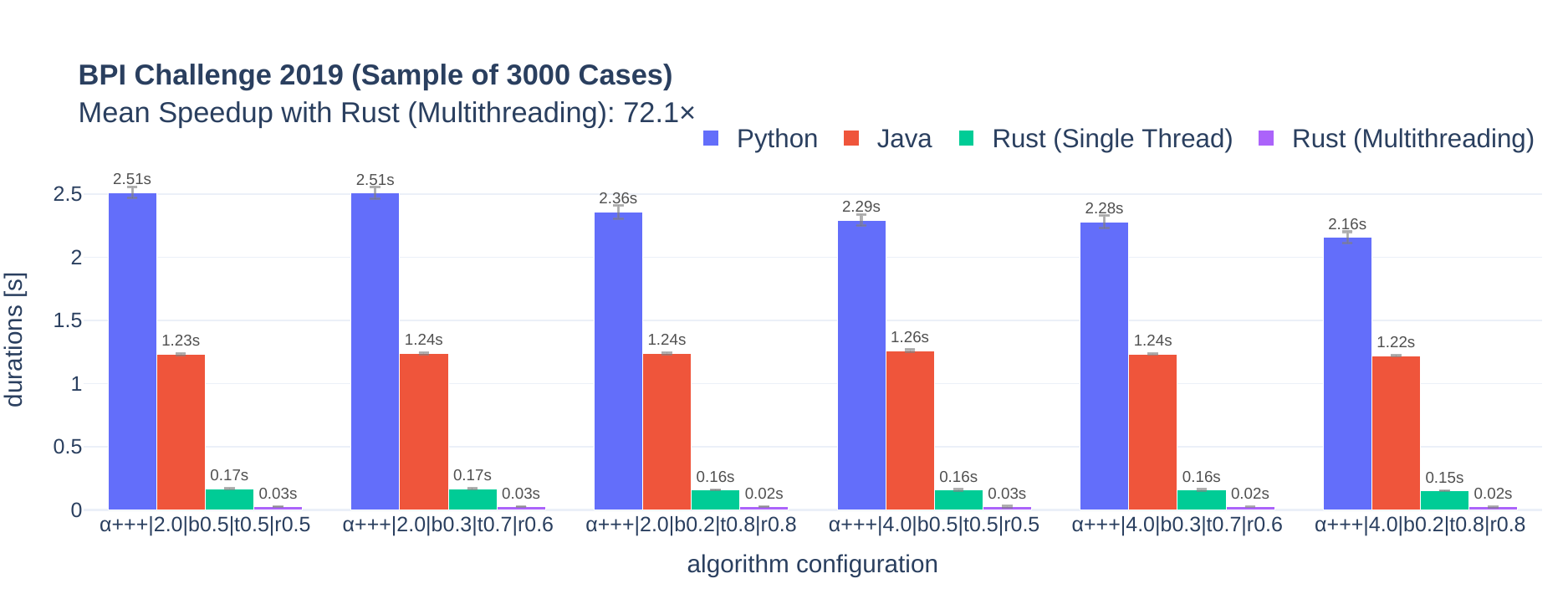}
    \includegraphics[width=0.95\textwidth]{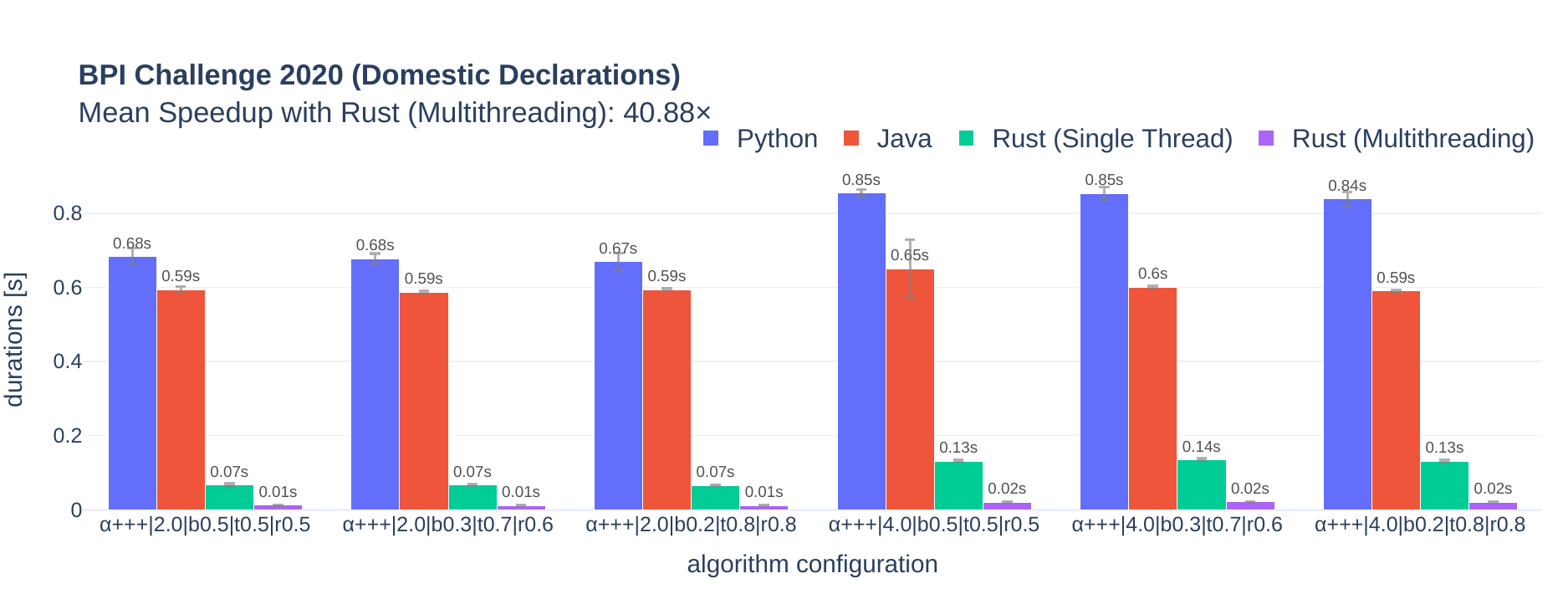}
    \includegraphics[width=0.95\textwidth]{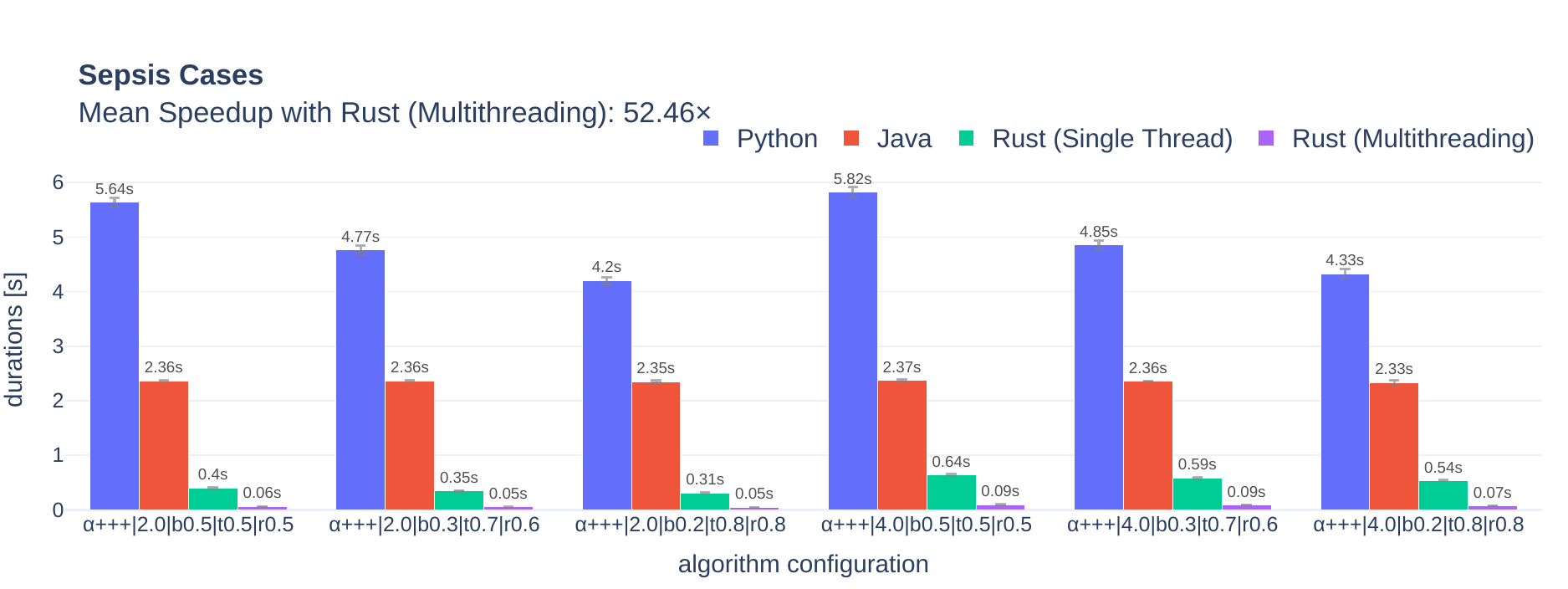}
    \includegraphics[width=0.95\textwidth]{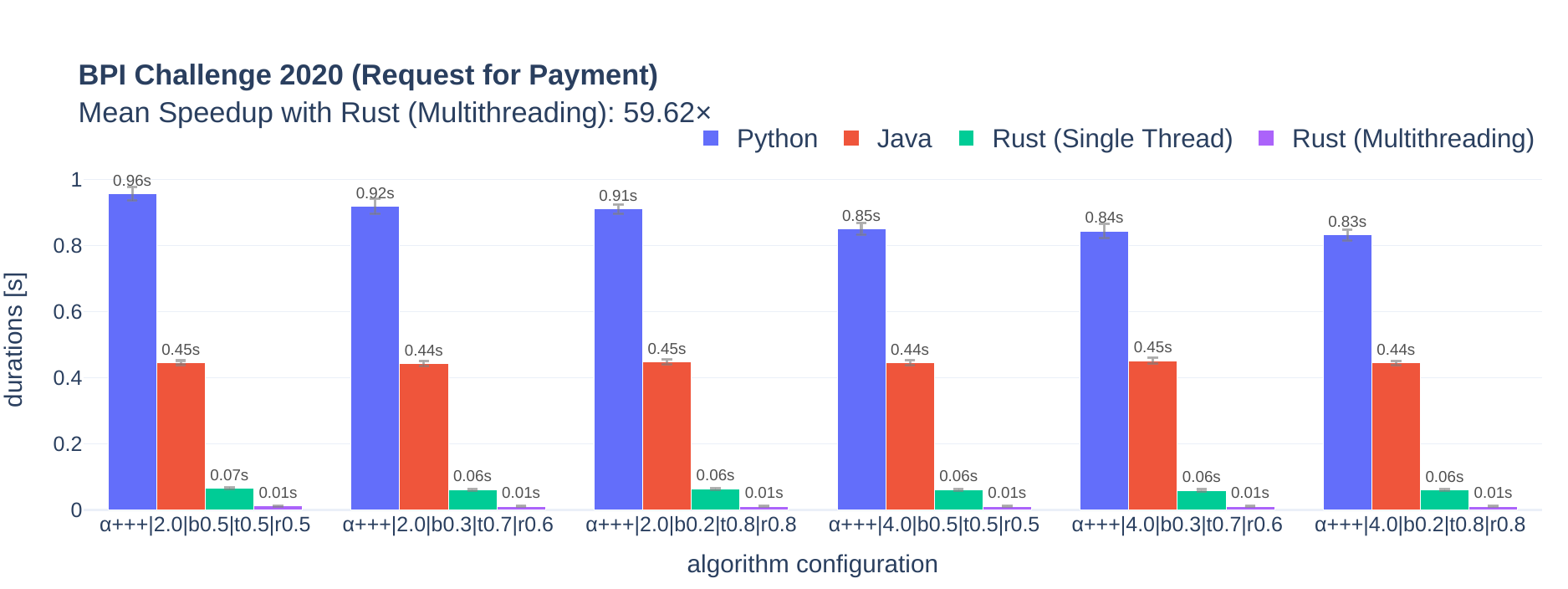}
    \caption{Performance comparison of the Alpha+++ discovery algorithm between the initial Python and Java implementation and the efficient Rust re-implementation across different event logs.}
    \label{fig:alphappp-performance}
\end{figure}

To evaluate the performance of the Alpha+++ Rust implementation, we measured the absolute discovery duration of the newly developed Rust implementation compared to the existing implementations in Java and Python (from \cite{alpha-revisit-2023}) across multiple commonly used event logs.
\autoref{fig:alphappp-performance} shows the execution times per event log and algorithm configuration for the different implementations.
Overall, the performance improvements of the Rust implementation are considerable.
The Rust implementation consistently achieves speedups of $50\times$ and more in many configurations compared to the Java and Python implementation.
Notably, the Rust implementation also makes use of parallelization, while the initial Java and Python implementations do not.
To measure the influence of multithreading in the Rust implementation, we additionally ran evaluations while restricting it to a single computing thread.
As expected, the single-threaded mode performs worse, but only by a factor of around $6$, compared to the multithreaded mode.
This indicates that the biggest performance differences between the Java, Python and Rust-based implementations are not caused by multithreading.

It is important to mention that we do not consider this evaluation to compare the maximum achievable performance in the different languages.
In particular, the Java and Python implementations were not separately optimized for performance after their initial creation.
Instead, we want to highlight the potential performance improvements that are achievable with a cross-platform Rust implementation in such scenarios.

\begin{figure}[h]
    \centering
    \includegraphics[width=0.7\textwidth]{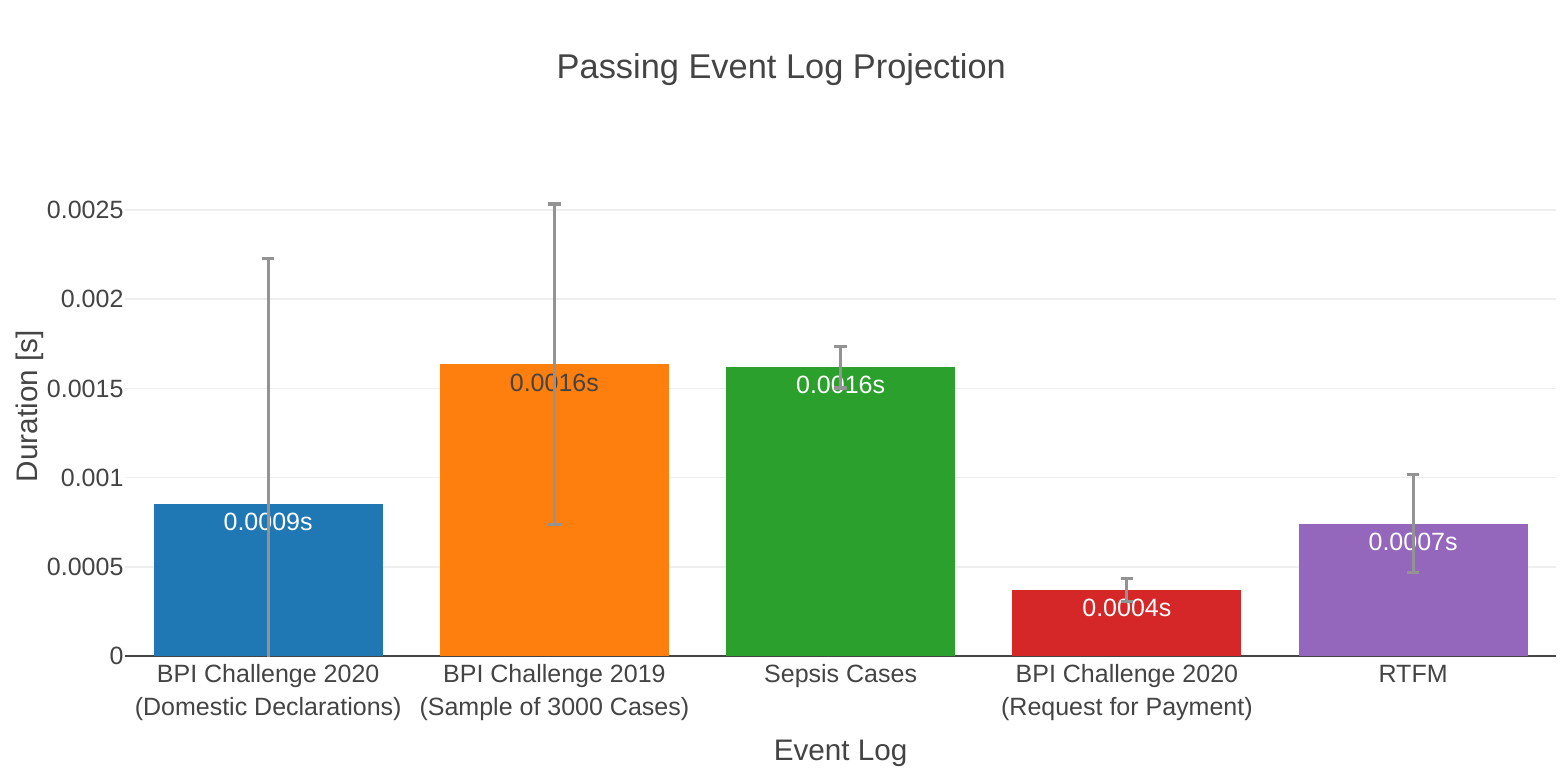}
    \includegraphics[width=0.7\textwidth]{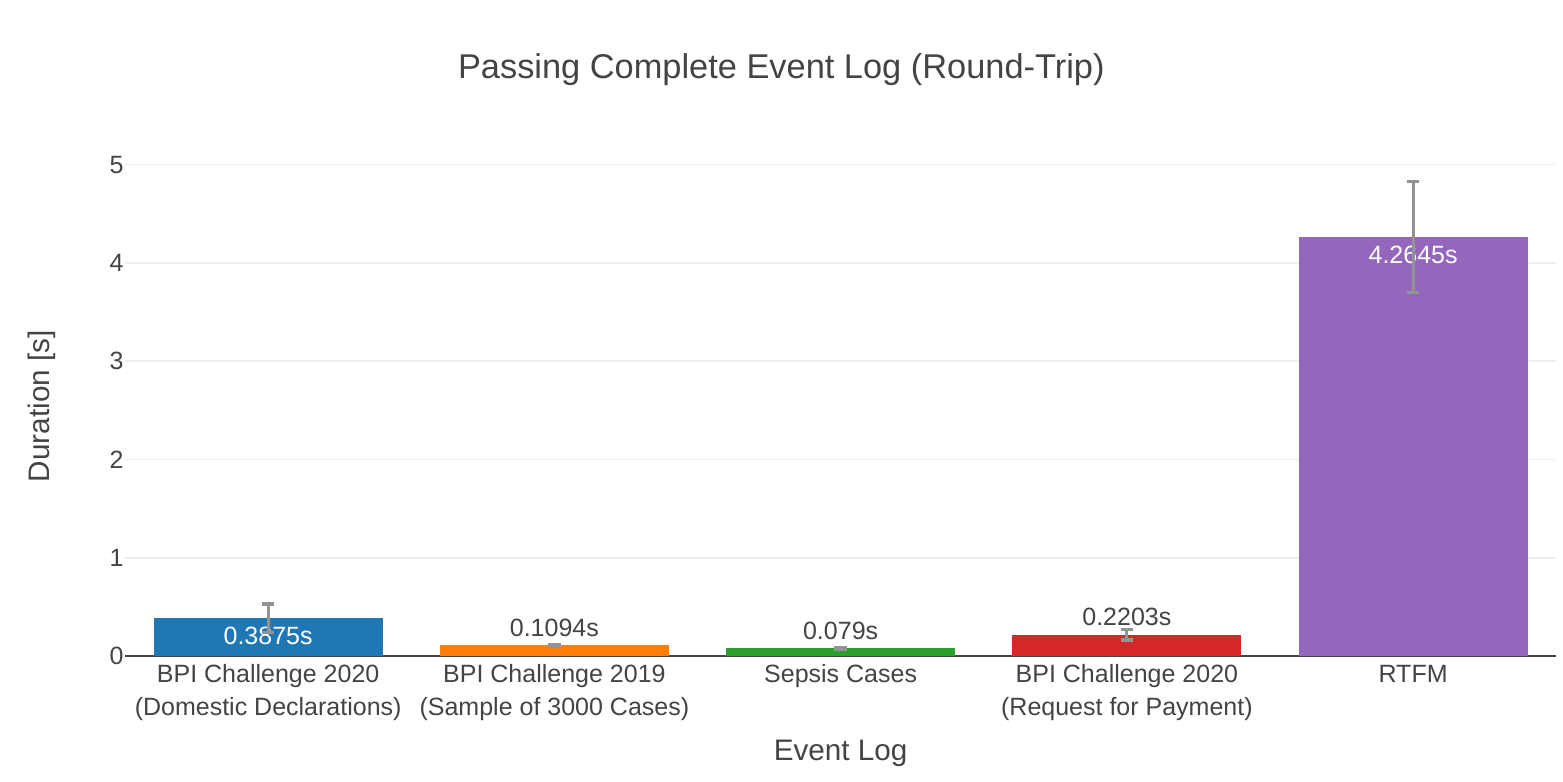}
    \caption{Duration of passing data from Java to Rust and back.
        On the top, only the activity projection of the event log is passed to Rust, while on the bottom, the complete event log is passed to Rust and back.
    }
    \label{fig:passing-data-java}
\end{figure}

In \autoref{fig:passing-data-java}, we additionally showcase measured durations for passing event log data from Java to Rust.
In the first scenario, the activity projection of an event log (as tuples of unique traces with their number of occurrences) is passed to Rust, which then computes the number of cases and returns that as a string.
The second scenario passes the complete event log to Rust, which then adds artificial start and end activities to each case and passes the modified event log back to Java.
These scenarios only involve very light computational work and thus mainly showcase the expected overhead of transferring event log data to or from Rust.
As expected, passing only the activity projection is significantly faster than passing the complete event log.
For activity projections, the measured durations are consistently under $2$ milliseconds.
Passing full event logs is also still fast (less than $0.5$ seconds) for most event logs, but it surpasses $4$ seconds for RTFM.
This is not only attributable to the larger number of cases and events in the RTFM event log, but also the relatively few unique variants of the RTFM log.
For the activity projection, only unique variants together with their count are transferred, which drastically compresses the data passed for RTFM.
This also explains the observed larger relative time difference between RTFM and the other logs when passing the complete event log instead of just the activity projection.

\clearpage
\section{Conclusion}
\label{sec:conclusion}
In this paper, we presented an approach for developing a shared Rust library with bindings to Java and Python, as well as possibly additional platforms and languages.
We introduced the field of process mining and described the current state of its software implementations, largely divided between the open-source projects PM4Py and ProM.
This division, in addition to duplicate implementation efforts and performance shortcomings, motivate our approach of developing one single shared performant Rust-based implementation.

We detailed the basic architecture and implementation of our approach.
Additionally, we provided a bootstrap template that can serve as a starting point for readers interested in developing their own implementation based on our approach\footnote{\url{https://github.com/aarkue/rust-bridge-template}}.

To demonstrate the feasibility and performance gain of our approach, we ported the Alpha+++ algorithm to Rust and developed an XES event log parser.
Our evaluation indicated a very significant performance speedup of up to $70\times$ for Alpha+++ and up to $5\times$ for our XES importer. 

For future work, it would be interesting to see evaluation of other resource-intensive algorithms ported to this Rust-based approach, for example other process discovery algorithms or the computation of alignments.
Additionally, it should be investigated how to create a framework around such implementations, also considering sharing code across different projects and implementing a plug-in system.

%
% ---- Bibliography ----
%
% BibTeX users should specify bibliography style 'splncs04'.
% References will then be sorted and formatted in the correct style.
%
\bibliographystyle{splncs04}
\bibliography{references}

\clearpage
\appendix
\section{Alpha+++ Runtime Evaluation Tables}
\label{appendix:alphappp-tables}
\begin{table}
    \centering
    \tiny
    \caption{Results comparing the runtime of different implementations of Alpha+++ on BPI Challenge 2020 (Domestic Declarations)}
    \begin{tabular}{lllccclll}
\hline
 Event Log                                                 & Alpha+++ Variant               & Implementation       &  Median  &  Mean   &   SD    \\
\hline
 \emph{BPI Challenge 2020}\\\emph{(Domestic Declarations)} &                                &                      &          &         &         \\
                                                           & $\alpha$+++|2.0|b0.5|t0.5|r0.5 &                      &          &         &         \\
                                                           &                                & Python               & 0.6761s  & 0.6761s & 0.0225s \\
                                                           &                                & Java                 & 0.5925s  & 0.5925s & 0.0080s \\
                                                           &                                & Rust (Single Thread) & 0.0675s  & 0.0675s & 0.0036s \\
                                                           &                                & Rust                 & 0.0120s  & 0.0120s & 0.0007s \\
                                                           & $\alpha$+++|2.0|b0.3|t0.7|r0.6 &                      &          &         &         \\
                                                           &                                & Python               & 0.6797s  & 0.6797s & 0.0137s \\
                                                           &                                & Java                 & 0.5875s  & 0.5875s & 0.0030s \\
                                                           &                                & Rust (Single Thread) & 0.0670s  & 0.0670s & 0.0026s \\
                                                           &                                & Rust                 & 0.0120s  & 0.0120s & 0.0005s \\
                                                           & $\alpha$+++|2.0|b0.2|t0.8|r0.7 &                      &          &         &         \\
                                                           &                                & Python               & 0.6736s  & 0.6736s & 0.0115s \\
                                                           &                                & Java                 & 0.6029s  & 0.6029s & 0.0094s \\
                                                           &                                & Rust (Single Thread) & 0.0650s  & 0.0650s & 0.0018s \\
                                                           &                                & Rust                 & 0.0120s  & 0.0120s & 0.0004s \\
                                                           & $\alpha$+++|2.0|b0.2|t0.8|r0.8 &                      &          &         &         \\
                                                           &                                & Python               & 0.6597s  & 0.6597s & 0.0235s \\
                                                           &                                & Java                 & 0.5944s  & 0.5944s & 0.0032s \\
                                                           &                                & Rust (Single Thread) & 0.0660s  & 0.0660s & 0.0016s \\
                                                           &                                & Rust                 & 0.0120s  & 0.0120s & 0.0008s \\
                                                           & $\alpha$+++|2.0|b0.1|t0.9|r0.9 &                      &          &         &         \\
                                                           &                                & Python               & 0.6407s  & 0.6407s & 0.0170s \\
                                                           &                                & Java                 & 0.5895s  & 0.5895s & 0.0027s \\
                                                           &                                & Rust (Single Thread) & 0.0655s  & 0.0655s & 0.0012s \\
                                                           &                                & Rust                 & 0.0120s  & 0.0120s & 0.0007s \\
                                                           & $\alpha$+++|4.0|b0.5|t0.5|r0.5 &                      &          &         &         \\
                                                           &                                & Python               & 0.8550s  & 0.8550s & 0.0106s \\
                                                           &                                & Java                 & 0.6149s  & 0.6149s & 0.0772s \\
                                                           &                                & Rust (Single Thread) & 0.1315s  & 0.1315s & 0.0021s \\
                                                           &                                & Rust                 & 0.0210s  & 0.0210s & 0.0007s \\
                                                           & $\alpha$+++|4.0|b0.3|t0.7|r0.6 &                      &          &         &         \\
                                                           &                                & Python               & 0.8521s  & 0.8521s & 0.0171s \\
                                                           &                                & Java                 & 0.6008s  & 0.6008s & 0.0027s \\
                                                           &                                & Rust (Single Thread) & 0.1350s  & 0.1350s & 0.0031s \\
                                                           &                                & Rust                 & 0.0210s  & 0.0210s & 0.0016s \\
                                                           & $\alpha$+++|4.0|b0.2|t0.8|r0.7 &                      &          &         &         \\
                                                           &                                & Python               & 0.8307s  & 0.8307s & 0.0188s \\
                                                           &                                & Java                 & 0.5930s  & 0.5930s & 0.0068s \\
                                                           &                                & Rust (Single Thread) & 0.1315s  & 0.1315s & 0.0058s \\
                                                           &                                & Rust                 & 0.0200s  & 0.0200s & 0.0009s \\
                                                           & $\alpha$+++|4.0|b0.2|t0.8|r0.8 &                      &          &         &         \\
                                                           &                                & Python               & 0.8393s  & 0.8393s & 0.0191s \\
                                                           &                                & Java                 & 0.5896s  & 0.5896s & 0.0036s \\
                                                           &                                & Rust (Single Thread) & 0.1310s  & 0.1310s & 0.0031s \\
                                                           &                                & Rust                 & 0.0205s  & 0.0205s & 0.0011s \\
                                                           & $\alpha$+++|4.0|b0.1|t0.9|r0.9 &                      &          &         &         \\
                                                           &                                & Python               & 0.8185s  & 0.8185s & 0.0166s \\
                                                           &                                & Java                 & 0.6025s  & 0.6025s & 0.0074s \\
                                                           &                                & Rust (Single Thread) & 0.1285s  & 0.1285s & 0.0041s \\
                                                           &                                & Rust                 & 0.0200s  & 0.0200s & 0.0016s \\
\hline
\end{tabular}
\end{table}

\begin{table}
    \centering
    \tiny
    \caption{Results comparing the runtime of different implementations of Alpha+++ on BPI Challenge 2020 (Request for Payment)}
    \begin{tabular}{lllccclll}
\hline
 Event Log                                               & Alpha+++ Variant               & Implementation       &  Median  &  Mean   &   SD    \\
\hline
 \emph{BPI Challenge 2020}\\\emph{(Request for Payment)} &                                &                      &          &         &         \\
                                                         & $\alpha$+++|2.0|b0.5|t0.5|r0.5 &                      &          &         &         \\
                                                         &                                & Python               & 0.9575s  & 0.9575s & 0.0207s \\
                                                         &                                & Java                 & 0.4485s  & 0.4485s & 0.0064s \\
                                                         &                                & Rust (Single Thread) & 0.0650s  & 0.0650s & 0.0035s \\
                                                         &                                & Rust                 & 0.0115s  & 0.0115s & 0.0007s \\
                                                         & $\alpha$+++|2.0|b0.3|t0.7|r0.6 &                      &          &         &         \\
                                                         &                                & Python               & 0.9210s  & 0.9210s & 0.0231s \\
                                                         &                                & Java                 & 0.4454s  & 0.4454s & 0.0079s \\
                                                         &                                & Rust (Single Thread) & 0.0610s  & 0.0610s & 0.0021s \\
                                                         &                                & Rust                 & 0.0110s  & 0.0110s & 0.0008s \\
                                                         & $\alpha$+++|2.0|b0.2|t0.8|r0.7 &                      &          &         &         \\
                                                         &                                & Python               & 0.9154s  & 0.9154s & 0.0104s \\
                                                         &                                & Java                 & 0.4484s  & 0.4484s & 0.0073s \\
                                                         &                                & Rust (Single Thread) & 0.0640s  & 0.0640s & 0.0022s \\
                                                         &                                & Rust                 & 0.0110s  & 0.0110s & 0.0003s \\
                                                         & $\alpha$+++|2.0|b0.2|t0.8|r0.8 &                      &          &         &         \\
                                                         &                                & Python               & 0.9113s  & 0.9113s & 0.0133s \\
                                                         &                                & Java                 & 0.4484s  & 0.4484s & 0.0070s \\
                                                         &                                & Rust (Single Thread) & 0.0630s  & 0.0630s & 0.0026s \\
                                                         &                                & Rust                 & 0.0110s  & 0.0110s & 0.0005s \\
                                                         & $\alpha$+++|2.0|b0.1|t0.9|r0.9 &                      &          &         &         \\
                                                         &                                & Python               & 0.9010s  & 0.9010s & 0.0181s \\
                                                         &                                & Java                 & 0.4457s  & 0.4457s & 0.0065s \\
                                                         &                                & Rust (Single Thread) & 0.0620s  & 0.0620s & 0.0021s \\
                                                         &                                & Rust                 & 0.0110s  & 0.0110s & 0.0006s \\
                                                         & $\alpha$+++|4.0|b0.5|t0.5|r0.5 &                      &          &         &         \\
                                                         &                                & Python               & 0.8591s  & 0.8591s & 0.0175s \\
                                                         &                                & Java                 & 0.4463s  & 0.4463s & 0.0070s \\
                                                         &                                & Rust (Single Thread) & 0.0605s  & 0.0605s & 0.0029s \\
                                                         &                                & Rust                 & 0.0110s  & 0.0110s & 0.0006s \\
                                                         & $\alpha$+++|4.0|b0.3|t0.7|r0.6 &                      &          &         &         \\
                                                         &                                & Python               & 0.8423s  & 0.8423s & 0.0207s \\
                                                         &                                & Java                 & 0.4527s  & 0.4527s & 0.0083s \\
                                                         &                                & Rust (Single Thread) & 0.0590s  & 0.0590s & 0.0029s \\
                                                         &                                & Rust                 & 0.0110s  & 0.0110s & 0.0006s \\
                                                         & $\alpha$+++|4.0|b0.2|t0.8|r0.7 &                      &          &         &         \\
                                                         &                                & Python               & 0.8335s  & 0.8335s & 0.0155s \\
                                                         &                                & Java                 & 0.4465s  & 0.4465s & 0.0046s \\
                                                         &                                & Rust (Single Thread) & 0.0615s  & 0.0615s & 0.0022s \\
                                                         &                                & Rust                 & 0.0110s  & 0.0110s & 0.0005s \\
                                                         & $\alpha$+++|4.0|b0.2|t0.8|r0.8 &                      &          &         &         \\
                                                         &                                & Python               & 0.8357s  & 0.8357s & 0.0161s \\
                                                         &                                & Java                 & 0.4475s  & 0.4475s & 0.0065s \\
                                                         &                                & Rust (Single Thread) & 0.0605s  & 0.0605s & 0.0024s \\
                                                         &                                & Rust                 & 0.0110s  & 0.0110s & 0.0009s \\
                                                         & $\alpha$+++|4.0|b0.1|t0.9|r0.9 &                      &          &         &         \\
                                                         &                                & Python               & 0.8202s  & 0.8202s & 0.0200s \\
                                                         &                                & Java                 & 0.4435s  & 0.4435s & 0.0068s \\
                                                         &                                & Rust (Single Thread) & 0.0580s  & 0.0580s & 0.0022s \\
                                                         &                                & Rust                 & 0.0110s  & 0.0110s & 0.0006s \\
\hline
\end{tabular}
\end{table}

\begin{table}
    \centering
    \tiny
    \caption{Results comparing the runtime of different implementations of Alpha+++ on BPI Challenge 2019 (Sample of 3000 Cases)}
    \begin{tabular}{lllccclll}
\hline
 Event Log                                                & Alpha+++ Variant               & Implementation       &  Median  &  Mean   &   SD    \\
\hline
 \emph{BPI Challenge 2019}\\\emph{(Sample of 3000 Cases)} &                                &                      &          &         &         \\
                                                          & $\alpha$+++|2.0|b0.5|t0.5|r0.5 &                      &          &         &         \\
                                                          &                                & Python               & 2.5053s  & 2.5053s & 0.0444s \\
                                                          &                                & Java                 & 1.2314s  & 1.2314s & 0.0072s \\
                                                          &                                & Rust (Single Thread) & 0.1660s  & 0.1660s & 0.0047s \\
                                                          &                                & Rust                 & 0.0260s  & 0.0260s & 0.0008s \\
                                                          & $\alpha$+++|2.0|b0.3|t0.7|r0.6 &                      &          &         &         \\
                                                          &                                & Python               & 2.5198s  & 2.5198s & 0.0475s \\
                                                          &                                & Java                 & 1.2391s  & 1.2391s & 0.0053s \\
                                                          &                                & Rust (Single Thread) & 0.1660s  & 0.1660s & 0.0056s \\
                                                          &                                & Rust                 & 0.0255s  & 0.0255s & 0.0007s \\
                                                          & $\alpha$+++|2.0|b0.2|t0.8|r0.7 &                      &          &         &         \\
                                                          &                                & Python               & 2.3436s  & 2.3436s & 0.0460s \\
                                                          &                                & Java                 & 1.2436s  & 1.2436s & 0.0038s \\
                                                          &                                & Rust (Single Thread) & 0.1610s  & 0.1610s & 0.0022s \\
                                                          &                                & Rust                 & 0.0240s  & 0.0240s & 0.0004s \\
                                                          & $\alpha$+++|2.0|b0.2|t0.8|r0.8 &                      &          &         &         \\
                                                          &                                & Python               & 2.3502s  & 2.3502s & 0.0529s \\
                                                          &                                & Java                 & 1.2367s  & 1.2367s & 0.0072s \\
                                                          &                                & Rust (Single Thread) & 0.1590s  & 0.1590s & 0.0022s \\
                                                          &                                & Rust                 & 0.0240s  & 0.0240s & 0.0008s \\
                                                          & $\alpha$+++|2.0|b0.1|t0.9|r0.9 &                      &          &         &         \\
                                                          &                                & Python               & 2.2289s  & 2.2289s & 0.0390s \\
                                                          &                                & Java                 & 1.2387s  & 1.2387s & 0.0023s \\
                                                          &                                & Rust (Single Thread) & 0.1580s  & 0.1580s & 0.0040s \\
                                                          &                                & Rust                 & 0.0235s  & 0.0235s & 0.0012s \\
                                                          & $\alpha$+++|4.0|b0.5|t0.5|r0.5 &                      &          &         &         \\
                                                          &                                & Python               & 2.2844s  & 2.2844s & 0.0442s \\
                                                          &                                & Java                 & 1.2575s  & 1.2575s & 0.0106s \\
                                                          &                                & Rust (Single Thread) & 0.1560s  & 0.1560s & 0.0044s \\
                                                          &                                & Rust                 & 0.0250s  & 0.0250s & 0.0041s \\
                                                          & $\alpha$+++|4.0|b0.3|t0.7|r0.6 &                      &          &         &         \\
                                                          &                                & Python               & 2.2720s  & 2.2720s & 0.0511s \\
                                                          &                                & Java                 & 1.2353s  & 1.2353s & 0.0024s \\
                                                          &                                & Rust (Single Thread) & 0.1565s  & 0.1565s & 0.0043s \\
                                                          &                                & Rust                 & 0.0245s  & 0.0245s & 0.0005s \\
                                                          & $\alpha$+++|4.0|b0.2|t0.8|r0.7 &                      &          &         &         \\
                                                          &                                & Python               & 2.1293s  & 2.1293s & 0.0423s \\
                                                          &                                & Java                 & 1.2281s  & 1.2281s & 0.0052s \\
                                                          &                                & Rust (Single Thread) & 0.1495s  & 0.1495s & 0.0054s \\
                                                          &                                & Rust                 & 0.0240s  & 0.0240s & 0.0007s \\
                                                          & $\alpha$+++|4.0|b0.2|t0.8|r0.8 &                      &          &         &         \\
                                                          &                                & Python               & 2.1709s  & 2.1709s & 0.0446s \\
                                                          &                                & Java                 & 1.2203s  & 1.2203s & 0.0051s \\
                                                          &                                & Rust (Single Thread) & 0.1505s  & 0.1505s & 0.0023s \\
                                                          &                                & Rust                 & 0.0230s  & 0.0230s & 0.0013s \\
                                                          & $\alpha$+++|4.0|b0.1|t0.9|r0.9 &                      &          &         &         \\
                                                          &                                & Python               & 2.0295s  & 2.0295s & 0.0331s \\
                                                          &                                & Java                 & 1.2401s  & 1.2401s & 0.0044s \\
                                                          &                                & Rust (Single Thread) & 0.1460s  & 0.1460s & 0.0011s \\
                                                          &                                & Rust                 & 0.0220s  & 0.0220s & 0.0006s \\
\hline
\end{tabular}
\end{table}

\begin{table}
    \centering
    \tiny
    \caption{Results comparing the runtime of different implementations of Alpha+++ on Sepsis Cases}
    \begin{tabular}{lllccclll}
\hline
 Event Log           & Alpha+++ Variant               & Implementation       &  Median  &  Mean   &   SD    \\
\hline
 \emph{Sepsis Cases} &                                &                      &          &         &         \\
                     & $\alpha$+++|2.0|b0.5|t0.5|r0.5 &                      &          &         &         \\
                     &                                & Python               & 5.6474s  & 5.6474s & 0.0758s \\
                     &                                & Java                 & 2.3630s  & 2.3630s & 0.0061s \\
                     &                                & Rust (Single Thread) & 0.4040s  & 0.4040s & 0.0032s \\
                     &                                & Rust                 & 0.0610s  & 0.0610s & 0.0017s \\
                     & $\alpha$+++|2.0|b0.3|t0.7|r0.6 &                      &          &         &         \\
                     &                                & Python               & 4.7470s  & 4.7470s & 0.0783s \\
                     &                                & Java                 & 2.3601s  & 2.3601s & 0.0096s \\
                     &                                & Rust (Single Thread) & 0.3490s  & 0.3490s & 0.0039s \\
                     &                                & Rust                 & 0.0530s  & 0.0530s & 0.0009s \\
                     & $\alpha$+++|2.0|b0.2|t0.8|r0.7 &                      &          &         &         \\
                     &                                & Python               & 4.2339s  & 4.2339s & 0.0722s \\
                     &                                & Java                 & 2.3656s  & 2.3656s & 0.0076s \\
                     &                                & Rust (Single Thread) & 0.3090s  & 0.3090s & 0.0025s \\
                     &                                & Rust                 & 0.0485s  & 0.0485s & 0.0026s \\
                     & $\alpha$+++|2.0|b0.2|t0.8|r0.8 &                      &          &         &         \\
                     &                                & Python               & 4.1869s  & 4.1869s & 0.0558s \\
                     &                                & Java                 & 2.3505s  & 2.3505s & 0.0161s \\
                     &                                & Rust (Single Thread) & 0.3090s  & 0.3090s & 0.0030s \\
                     &                                & Rust                 & 0.0480s  & 0.0480s & 0.0008s \\
                     & $\alpha$+++|2.0|b0.1|t0.9|r0.9 &                      &          &         &         \\
                     &                                & Python               & 3.4777s  & 3.4777s & 0.0848s \\
                     &                                & Java                 & 2.3572s  & 2.3572s & 0.0073s \\
                     &                                & Rust (Single Thread) & 0.2665s  & 0.2665s & 0.0025s \\
                     &                                & Rust                 & 0.0420s  & 0.0420s & 0.0022s \\
                     & $\alpha$+++|4.0|b0.5|t0.5|r0.5 &                      &          &         &         \\
                     &                                & Python               & 5.8432s  & 5.8432s & 0.0987s \\
                     &                                & Java                 & 2.3702s  & 2.3702s & 0.0174s \\
                     &                                & Rust (Single Thread) & 0.6415s  & 0.6415s & 0.0122s \\
                     &                                & Rust                 & 0.0910s  & 0.0910s & 0.0049s \\
                     & $\alpha$+++|4.0|b0.3|t0.7|r0.6 &                      &          &         &         \\
                     &                                & Python               & 4.8226s  & 4.8226s & 0.0836s \\
                     &                                & Java                 & 2.3603s  & 2.3603s & 0.0041s \\
                     &                                & Rust (Single Thread) & 0.5855s  & 0.5855s & 0.0050s \\
                     &                                & Rust                 & 0.0870s  & 0.0870s & 0.0039s \\
                     & $\alpha$+++|4.0|b0.2|t0.8|r0.7 &                      &          &         &         \\
                     &                                & Python               & 4.3592s  & 4.3592s & 0.0485s \\
                     &                                & Java                 & 2.3580s  & 2.3580s & 0.0041s \\
                     &                                & Rust (Single Thread) & 0.5360s  & 0.5360s & 0.0085s \\
                     &                                & Rust                 & 0.0750s  & 0.0750s & 0.0035s \\
                     & $\alpha$+++|4.0|b0.2|t0.8|r0.8 &                      &          &         &         \\
                     &                                & Python               & 4.3240s  & 4.3240s & 0.0844s \\
                     &                                & Java                 & 2.3209s  & 2.3209s & 0.0432s \\
                     &                                & Rust (Single Thread) & 0.5365s  & 0.5365s & 0.0073s \\
                     &                                & Rust                 & 0.0740s  & 0.0740s & 0.0019s \\
                     & $\alpha$+++|4.0|b0.1|t0.9|r0.9 &                      &          &         &         \\
                     &                                & Python               & 3.7477s  & 3.7477s & 0.0651s \\
                     &                                & Java                 & 2.3627s  & 2.3627s & 0.0233s \\
                     &                                & Rust (Single Thread) & 0.4835s  & 0.4835s & 0.0051s \\
                     &                                & Rust                 & 0.0670s  & 0.0670s & 0.0010s \\
\hline
\end{tabular}
\end{table}

\begin{table}
    \centering
    \tiny
    \caption{Results comparing the runtime of different implementations of Alpha+++ on RTFM}
    \begin{tabular}{lllccclll}
\hline
 Event Log   & Alpha+++ Variant               & Implementation       &  Median  &  Mean   &   SD    \\
\hline
 \emph{RTFM} &                                &                      &          &         &         \\
             & $\alpha$+++|2.0|b0.5|t0.5|r0.5 &                      &          &         &         \\
             &                                & Python               & 0.1702s  & 0.1702s & 0.0032s \\
             &                                & Java                 & 0.0904s  & 0.0904s & 0.0010s \\
             &                                & Rust (Single Thread) & 0.0350s  & 0.0350s & 0.0011s \\
             &                                & Rust                 & 0.0100s  & 0.0100s & 0.0006s \\
             & $\alpha$+++|2.0|b0.3|t0.7|r0.6 &                      &          &         &         \\
             &                                & Python               & 0.1603s  & 0.1603s & 0.0033s \\
             &                                & Java                 & 0.0899s  & 0.0899s & 0.0014s \\
             &                                & Rust (Single Thread) & 0.0340s  & 0.0340s & 0.0014s \\
             &                                & Rust                 & 0.0100s  & 0.0100s & 0.0005s \\
             & $\alpha$+++|2.0|b0.2|t0.8|r0.7 &                      &          &         &         \\
             &                                & Python               & 0.1466s  & 0.1466s & 0.0029s \\
             &                                & Java                 & 0.0910s  & 0.0910s & 0.0017s \\
             &                                & Rust (Single Thread) & 0.0340s  & 0.0340s & 0.0013s \\
             &                                & Rust                 & 0.0090s  & 0.0090s & 0.0007s \\
             & $\alpha$+++|2.0|b0.2|t0.8|r0.8 &                      &          &         &         \\
             &                                & Python               & 0.1490s  & 0.1490s & 0.0026s \\
             &                                & Java                 & 0.0909s  & 0.0909s & 0.0014s \\
             &                                & Rust (Single Thread) & 0.0340s  & 0.0340s & 0.0010s \\
             &                                & Rust                 & 0.0090s  & 0.0090s & 0.0004s \\
             & $\alpha$+++|2.0|b0.1|t0.9|r0.9 &                      &          &         &         \\
             &                                & Python               & 0.1412s  & 0.1412s & 0.0021s \\
             &                                & Java                 & 0.0907s  & 0.0907s & 0.0013s \\
             &                                & Rust (Single Thread) & 0.0340s  & 0.0340s & 0.0034s \\
             &                                & Rust                 & 0.0090s  & 0.0090s & 0.0007s \\
             & $\alpha$+++|4.0|b0.5|t0.5|r0.5 &                      &          &         &         \\
             &                                & Python               & 0.1825s  & 0.1825s & 0.0038s \\
             &                                & Java                 & 0.0913s  & 0.0913s & 0.0405s \\
             &                                & Rust (Single Thread) & 0.0365s  & 0.0365s & 0.0014s \\
             &                                & Rust                 & 0.0110s  & 0.0110s & 0.0005s \\
             & $\alpha$+++|4.0|b0.3|t0.7|r0.6 &                      &          &         &         \\
             &                                & Python               & 0.1737s  & 0.1737s & 0.0037s \\
             &                                & Java                 & 0.0905s  & 0.0905s & 0.0021s \\
             &                                & Rust (Single Thread) & 0.0350s  & 0.0350s & 0.0009s \\
             &                                & Rust                 & 0.0100s  & 0.0100s & 0.0009s \\
             & $\alpha$+++|4.0|b0.2|t0.8|r0.7 &                      &          &         &         \\
             &                                & Python               & 0.1550s  & 0.1550s & 0.0022s \\
             &                                & Java                 & 0.0901s  & 0.0901s & 0.0020s \\
             &                                & Rust (Single Thread) & 0.0350s  & 0.0350s & 0.0015s \\
             &                                & Rust                 & 0.0100s  & 0.0100s & 0.0006s \\
             & $\alpha$+++|4.0|b0.2|t0.8|r0.8 &                      &          &         &         \\
             &                                & Python               & 0.1554s  & 0.1554s & 0.0027s \\
             &                                & Java                 & 0.0900s  & 0.0900s & 0.0023s \\
             &                                & Rust (Single Thread) & 0.0350s  & 0.0350s & 0.0027s \\
             &                                & Rust                 & 0.0090s  & 0.0090s & 0.0005s \\
             & $\alpha$+++|4.0|b0.1|t0.9|r0.9 &                      &          &         &         \\
             &                                & Python               & 0.1479s  & 0.1479s & 0.0022s \\
             &                                & Java                 & 0.0901s  & 0.0901s & 0.0012s \\
             &                                & Rust (Single Thread) & 0.0340s  & 0.0340s & 0.0017s \\
             &                                & Rust                 & 0.0090s  & 0.0090s & 0.0005s \\
\hline
\end{tabular}
\end{table}

\end{document}